\documentclass[12pt,aps,onecolumn,superscriptaddress]{revtex4}

\usepackage{graphicx}
\usepackage{amsmath}
\usepackage{hyperref}

\usepackage{ifthen}
\newboolean{linenum}

\begin{document}

\title{\boldmath {Measurements of the neutron timelike electromagnetic
form factor with the SND detector} }
\ifthenelse{\boolean{linenum}}{
\linenumbers}{}

\newcommand{\binp}{\affiliation{Budker Institute of Nuclear Physics,
SB RAS, Novosibirsk 630090, Russia}}
\newcommand{\nsu}{\affiliation{Novosibirsk State University,
Novosibirsk 630090, Russia}}

\author{M.~N.~Achasov} \binp\nsu
\author{A.~Yu.~Barnyakov} \binp\nsu
\author{E.~V.~Bedarev} \binp\nsu\
\author{K.~I.~Beloborodov} \binp\nsu
\author{A.~V.~Berdyugin} \binp\nsu
\author{D.~E.~Berkaev}\binp
\author{A.~G.~Bogdanchikov}\binp
\author{A.~A.~Botov}\binp
\author{T.~V.~Dimova}\binp\nsu
\author{V.~P.~Druzhinin}\binp\nsu
\author{V.~N.~Zhabin}\binp\nsu
\author{Yu.~M.~Zharinov}\binp
\author{L.~V.~Kardapoltsev}\binp\nsu
\author{A.~S.~Kasaev}\binp
\author{D.~P.~Kovrizhin} \binp
\author{I.~A.~Koop}\binp\nsu
\author{A.~A.~Korol}\binp\nsu
\author{A.~S.~Kupich} \binp\nsu
\author{A.~P.~Kryukov} \binp
\author{A.~P.~Lysenko} \binp
\author{N.~A.~Melnikova} \binp\nsu
\author{N.~Yu.~Muchnoi} \binp\nsu
\author{A.~E.~Obrazovsky} \binp
\author{E.~V.~Pakhtusova} \binp
\author{K.~V.~Pugachev} \binp\nsu
\author{S.~A.~Rastigeev} \binp
\author{Yu.~A.~Rogovsky} \binp\nsu
\author{S.~I.~Serednyakov} \binp\nsu
\author{Z.~K.~Silagadze} \binp\nsu
\author{I.~K.~Surin} \binp
\author{Yu.~V.~Usov} \binp
\author{A.~G.~Kharlamov}\binp\nsu
\author{Yu.~M.~Shatunov} \binp
\author{D.~A.~Shtol} \binp

\date{ }

\begin{abstract}
The results of the measurement of the $e^+e^-\to n\bar{n}$ cross section
and effective neutron timelike form factor are presented.
The data taking was carried out in 2020--2021 at the VEPP-2000
$e^+e^-$ collider in the center-of-mass energy range from 1891 to 2007
MeV.
The general purpose nonmagnetic detector SND is used to detect
neutron-antineutrons events. The selection of $n\bar{n}$ events is
performed using the time-of-flight  technique. The measured cross
section is 0.4--0.6 nb.  The neutron form factor in the energy range
under study varies from 0.3 to 0.2.
\end{abstract}
\maketitle

\section*{Introduction\label{sec:intro}}
The internal structure of nucleons is described by electromagnetic
formfactors. In the timelike region they are measured using the process of 
$e^+e^-$ annihilation to nucleon-antinucleon pairs. The $e^+e^-\to n\bar{n}$ cross 
section  depends on two formfactors - electric $G_E$ and magnetic $G_M$ :
\begin{eqnarray}
\frac{d\sigma}{d\Omega}&=&\frac{\alpha^{2}\beta}{4s}
\bigg[ |G_M(s)|^{2}(1+\cos^2\theta)\nonumber\\
&+&\frac{1}{\gamma^2}|G_E(s)|^{2}\sin^2\theta
\bigg]
\label{eqB1}
\end{eqnarray}
where $\alpha$ is the fine structure constant, 
$s=4E_b^2=E^2$, where $E_b$ is the beam  energy and $E$ is the
center-of-mass (c.m.)  energy,  $\beta = \sqrt{1-4m_n^2/s}$, $\gamma
= E_b/m_n$,  $m_n$ is the neutron
mass   and $\theta$ is the antineutron production polar angle.
The total cross section has the following form:
\begin{equation}
\sigma(s) =
\frac{4\pi\alpha^{2}\beta}{3s}(1+\frac{1}{2\gamma^2})|F(s)|^2,
\label{eqB2}
\end{equation}
where   the effective form factor  $F(s)$ is introduced:  
\begin{equation}
|F(s)|^2=\frac{2\gamma^2|G_M(s)|^2+|G_E(s)|^2}{2\gamma^2 +1 }.
\label{eqB3}
\end{equation}
The $|G_E/G_M|$ ratio can be extracted
from the analysis of the measured $\cos\theta$ distribution in 
 Eq.~(\ref{eqB1}). At the threshold  $|G_{E}| = |G_{M}|$.

  The latest results on the neutron form factor near the
threshold were obtained in  experiments at the VEPP-2000 $e^+e^-$ collider with 
the SND detector ~\cite{SNND}. The same work provides a list of
previous measurements. At the energy above 2 GeV new data have
been obtained by the BESIII ~\cite{BES}. In this work the recent SND results  
on  the $e^+e^-\to n\bar{n}$ 
cross section and the neutron timelike formfactor with 4 times higher
integrated luminosity than in previous measurement ~\cite{SNND},   
are presented.  

\begin{figure*}
\centering
\includegraphics [width = 0.7\textwidth]{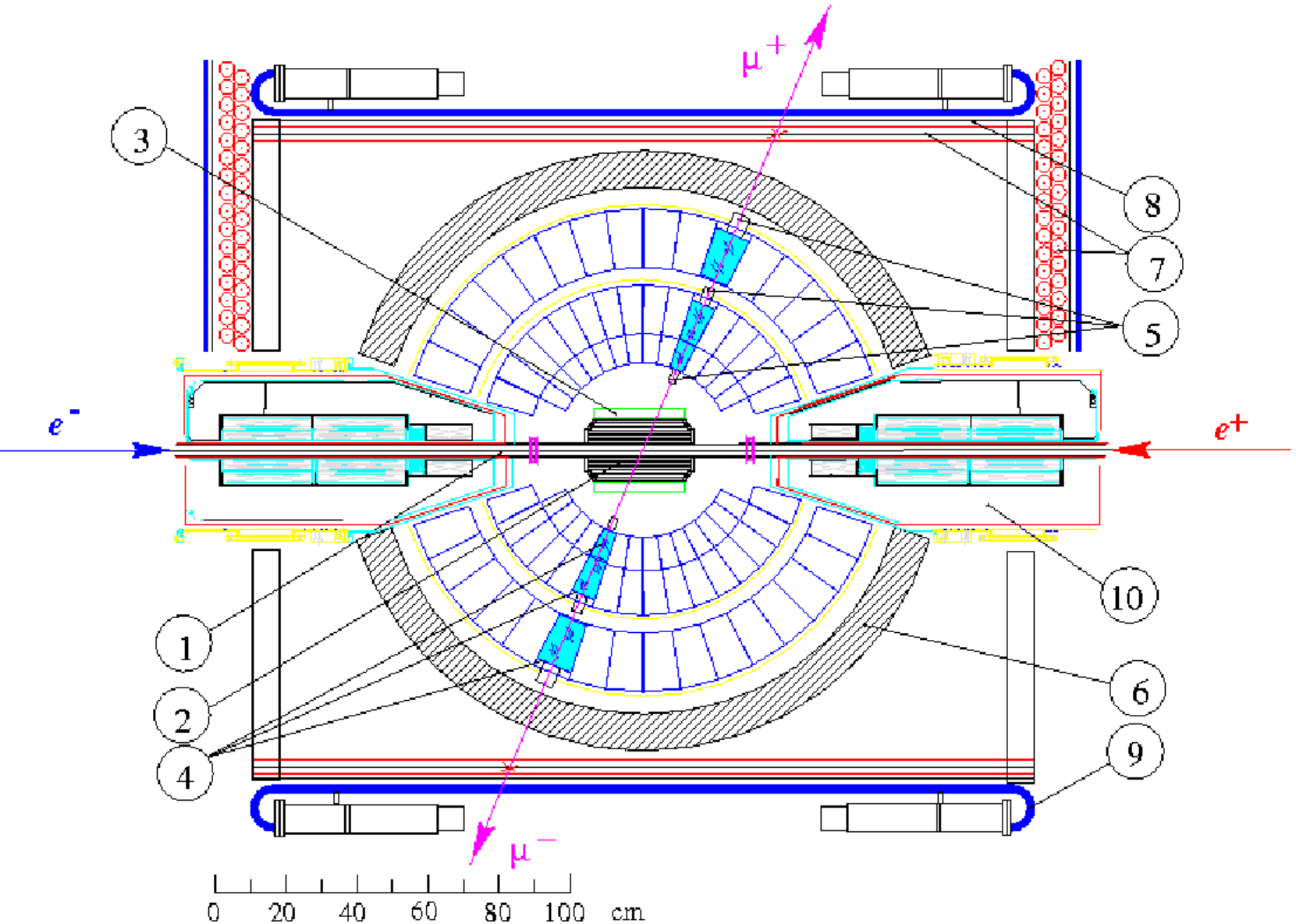}
\caption{SND detector, section along the beams: (1) beam pipe,
(2) tracking system, (3) aerogel Cherenkov counters, (4) NaI (Tl)
crystals, (5) vacuum phototriodes, (6) iron absorber, (7) proportional
tubes,
(8) iron absorber, (9) scintillation counters, (10) VEPP-2000 focusing
solenoids.}
\label{fig:sndt}
\end{figure*}

\section{Collider, detector, experiment\label{sec:Exper}}
VEPP-2000 is $e^+e^-$ collider~\cite{VEPP2k} operating in the energy range 
from the hadron threshold ($E$=280 MeV) up to 2 GeV. The collider
luminosity above the nucleon threshold at 1.87 GeV is of order of $5\times
10^{31}$~cm$^{-2}$s$^{-1}$. There are two
collider detectors at VEPP-2000: SND and CMD-3.

    SND (Spherical Neutral Detector) ~\cite{SNDet1}  is a non-magnetic detector,
including a  tracking system, a spherical NaI(Tl) electromagnetic calorimeter
(EMC) and a muon detector (Fig.\ref{fig:sndt}). 
The EMC is the main part of the SND used in the $n\bar{n}$ analysis. 
The thickness of EMC is 34.7 cm (13.4 radiation length).
The antineutron annihilation length in NaI(Tl)   
varies with energy from several cm close to the $n\bar{n}$ threshold 
to $\sim$15 cm at  the maximum available energy ~\cite{Annih}, 
so nearly all produced antineutrons  are absorbed in the detector. 

  The EMC is used to measure the event arrival time.  
Starting from 2019 a  system of flash ADC modules ~\cite{Timr},
measuring the signal waveform, is 
installed  on each of the 1640 EMC counters. When
fitting the flash ADC output waveform, the time and amplitude 
of the signal in the counters are
calculated. The event time is calculated as the energy weighted average
time. The time resolution obtained with $e^+e^-\to \gamma\gamma$
events is about 0.8 ns.

This article presents the analysis results of data with the integrated
luminosity of $80$ pb$^{-1}$,  collected in the energy
range $1.89$ - $2.0$  GeV in 8 energy points.  

\section{Selection of $n\bar{n}$ events \label{sec:EvSelect}}
Antineutron from the $n\bar{n}$ pair in most cases annihilates,
producing pions, nucleons, photons and other particles, 
which deposit up to 2 GeV in EMC. The neutron from the $n\bar{n}$ pair
release a small signal in EMC, which poorly visible against the 
background of a strong $\bar{n}$ annihilation signal, so it is not 
taken into account. The $n\bar{n}$ events are reconsructed as
multiphoton events.

   Main features of   $n\bar{n}$ events  are
absence of charged tracks and photons from the collision region and a
strong imbalance in the event momentum. To create the $n\bar{n}$
selection conditions we consider the sources of the background
including  the cosmic  background,  the background from $e^+e^-$ 
annihilation processes and from the electron and positron beams in the
collider.

   Based on these specific features of the $e^+e^-\to n\bar{n}$ process,
selection conditions were divided into three groups.

 In the first group the  conditions are collected that suppress the
background from the $e^+e^-$ annihilation events.
These include the condition of no charged tracks  in an event, 
the limit on the total event momentum (p/2$E_b >$0.4), 
and a limitation on the transverse shower profile in EMC ~\cite{xi2gam}, 
which must be wider than that from  the photons from the collision region. 

   In the second group the selection conditions should suppress  the 
cosmic   background.  Here the veto of muon system is included and 
special conditions, analyzing  the energy deposition shape in EMC and 
removing cosmic events passing through the muon veto ~\cite{SNND}. 
Basically, these are cosmic showers in EMC.

  The third group of selection cuts contains
the restriction on the total energy deposition in EMC ---  $E_{dep} > E_b$.
Such a restriction almost completely suppresses the beam background,
although the detection efficiency also decreases by about  20\%.

    The listed selection conditions are similar to those described in our
recent work ~\cite{SNND}. The only difference is that there is no
limitation  on the energy in the EMC third layer. This slightly
increased the detection efficiency, although it did increase the
cosmic background. After imposing the described selection conditions,
we have about 400 events/$pb$ left for further analysis.
\begin{figure*}
\includegraphics[width=0.48\textwidth]{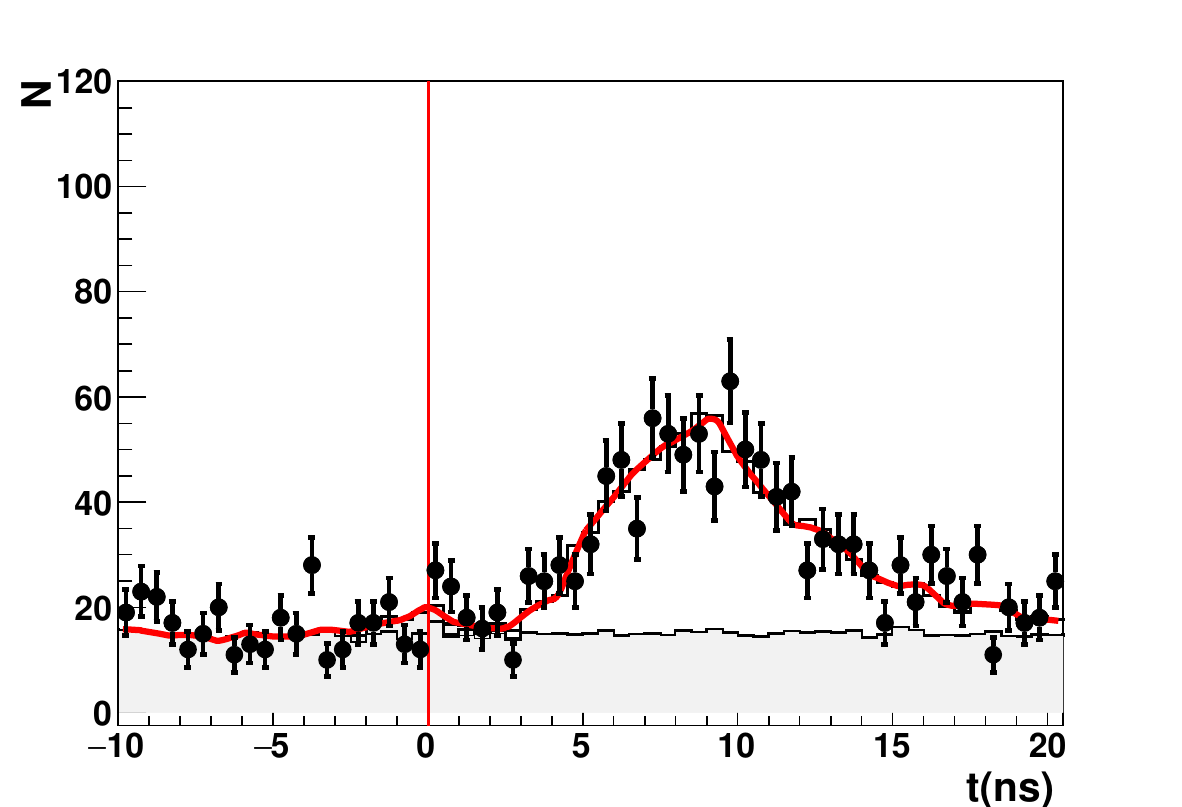} \hfill
\includegraphics[width=0.48\textwidth]{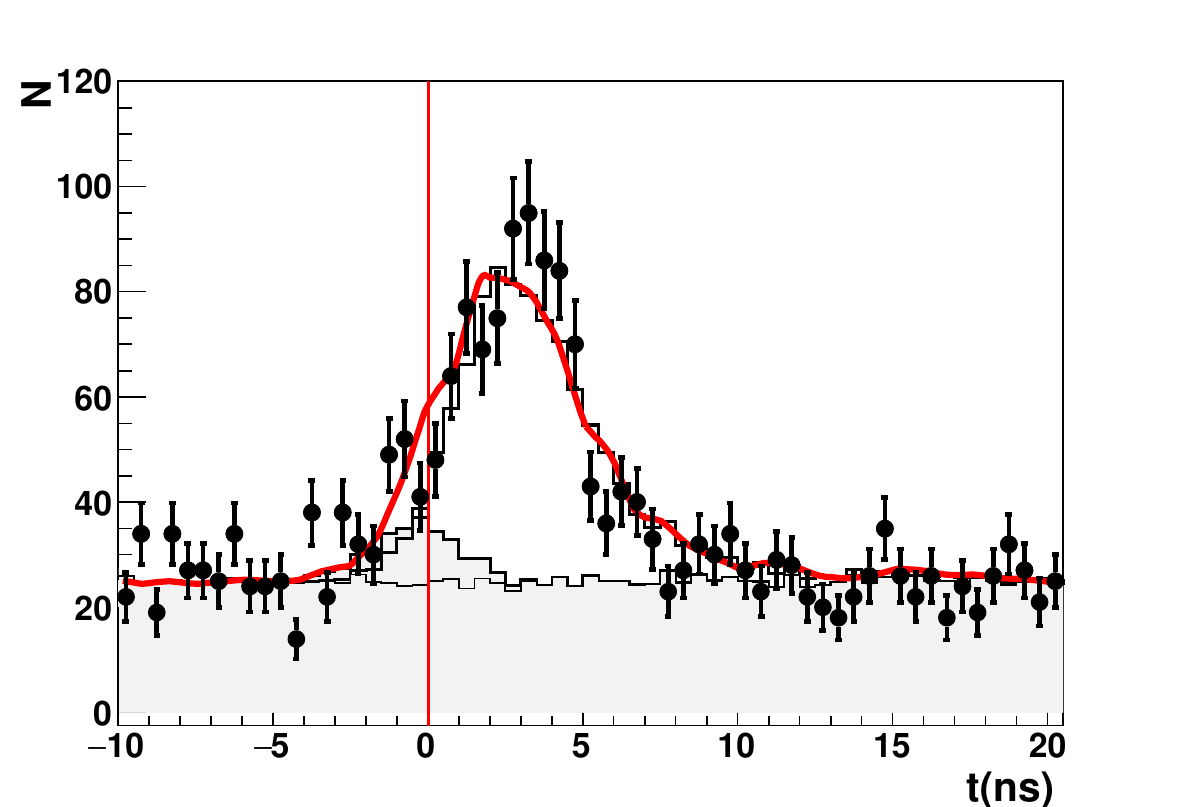}
\caption {The time distribution of  selected data events
(points with error bars) at $E_b=945$ MeV (left panel)
and at $E_b=980$ MeV (right panel). The position $t=0$ corresponds to
the moment of  beams collision. The wide peak to the right is 
the contribution of $n\bar{n}$ events. The light-shaded
histogram shows the cosmic background (uniform in time) and beam
background (peak at $t=0$). The solid line is the result of the fit. 
\label{fig:hstim}}
\end{figure*}
\section{Getting  the number of $n\bar{n}$ events 
\label{sec:Tim19}}
The time spectra for selected data events are shown in
Fig.~\ref{fig:hstim}. Zero time corresponds to events 
in the moment of beam collision. Three main components
are distinguished in the time spectrum in the figures shown: 
a beam background at t=0, a
cosmic background uniform in time,  and a delayed 
signal from $n\bar{n}$ events, wide in time. Respectively, the measured
time spectra are fitted by the sum of these three components in the
following form :
\begin{equation}
F(t)=N_{n\bar{n}}H_{n\bar{n}}(t)+N_{\rm csm}H_{\rm csm}(t)+N_{\rm
bkg}H_{\rm bkg}(t),
\label{eq17}
\end{equation}
where  $H_{n\bar{n}}$, $H_{\rm csm}$ and $H_{\rm bkg}$ are 
normalized histograms, describing  time spectra for the $n\bar{n}$ signal,
cosmic  and beam + physical  background, respectively.
$N_{n\bar{n}}$, $N_{\rm csm}$, and $N_{\rm bkg}$ are the corresponding  event
numbers, obtained from the fit. The shape of the beam+physical
background time spectrum  
$H_{\rm bkg}$ is measured at the energies below the $n\bar{n}$ threshold. 
The cosmic  time spectrum $H_{\rm csm}$ is measured with the lower EMC
threshold $0.9\cdot E_b$  in coincidence with the muon system signal.
The $H_{n\bar{n}}$ spectrum is calculated by the MC simulation the
$e^+e^-\to n\bar{n}$ process.

 Comparison of time spectra in data and MC gives a wider data
time distribution for both $e^+e^-\to \gamma\gamma$ and $e^+e^-\to
n\bar{n}$ events. For the $e^+e^-\to \gamma\gamma$ this is due to the
finite time resolution of our timing system ~\cite{Timr}, 
which is not adequately simulated.  So we
convolve the MC time spectrum with a Gaussian with $\sigma_{\gamma\gamma}=$
0.8 ns. For $e^+e^-\to n\bar{n}$ events the covolution is done with 
$\sigma_{nn} = 1.5$--2  ns depending on the energy.

    Moreover, in addition to the above, we  correct the  MC ${n\bar{n}}$ 
time spectrum, since the shape of MC time spectrum  $H_{n\bar{n}}$ does not 
describe data well.  This discrepancy is explained by the incorrect 
relationship between the processes of antineutron annihilation and 
scattering in  MC, as well as by the incorrect  description of the 
annihilation products. 
To modify the MC time spectrum,
separate MC time spectra were plotted for the cases of the first
$\bar{n}$  interaction of scattering ($H_{n\bar{n}}^{s}$)  and 
annihilation ($H_{n\bar{n}}^{a}$). The share of annihilation
events in MC was about 33\%.
The annihilation gives the time spectrum close to the  
exponential while the scattering has delayed and more wide time spectrum 
with the non exponential shape.   
The $H_{n\bar{n}}$ spectrum (Eq.\ref{eq17}) was taken in fit
as a linear sum of two spectra described above: 
$H_{n\bar{n}}=\alpha H_{n\bar{n}}^{a}+ (1-\alpha)H_{n\bar{n}}^{s}$ .
The value $\alpha$ (the  share of annihilation events) was the fit parameter. 
As a result of the fit this parameter turned out to be greater 
than in MC -- $\simeq$60\% and accordingly the proportion of scattering 
fell to $\simeq$40\%. 
As can be seen in Fig.\ref{fig:hstim}, 
the modified MC  time spectrum describes the data well.

 The visible cross section  $\sigma_{bg}$  of the beam+physical background, 
obtained during fitting, is about 7 pb and does not significantly
depend on the beam energy. The main contribution into $\sigma_{bg}$ comes from
the processes with neutral kaons in the final state: $e^+e^-\to
K_SK_L\pi^0$, $K_SK_L\eta$ and similar other.  The measured residual
cosmic background  rate has the  intensity $\sim$0.01 Hz, which
corresponds to the suppression of the number of cosmic events, that
have pass the hardware selection in the detector electronics,
approximately by $2\times 10^4$ times.

\begin{table*}
\centering
\caption{The beam energy ($E_b$), integrated luminosity ($L$),
number of selected $n\bar{n}$ events ($N_{n\bar{n}}$), the factor taking into
account radiative corrections and energy spread ($1+\delta$), corrected detection
efficiency ($\varepsilon$), measured $e^+e^-\to n\bar{n}$ cross
section $\sigma$, and neutron effective form factor ($F_n$). 
The quoted errors for $N$, $\sigma$ are statistical and 
systematic. For the detection efficiency, the systematic uncertainty
is quoted. For $F_n$, the combined statistical and systematic uncertainty is
listed.  
\label{tab:crsect}}
\begin{tabular}{cccccccc}
N & $E_b$(MeV) & $L$(pb)& $N_{n\bar{n}}$ & $1+\delta$ &$\varepsilon$
&$\sigma$(nb) & $F_n$\\
\hline
1 & 945.5  & 8.54 & $676\pm37$ & 0.746 & $0.253 \pm 0.021$ &
$0.420\pm0.023\pm0.036$ &$0.322\pm0.016$ \\
2 & 950.3  & 8.86 & $834\pm37$ & 0.787 & $0.246 \pm0.015$ &
$0.485\pm0.022\pm0.031$ &$0.301\pm0.012$ \\
3 & 960.3 & 8.33 & $767\pm35$ & 0.840 & $0.217\pm0.013$ &
$0.506\pm0.023\pm0.032$ &$0.266\pm0.010$ \\
4 & 970.8  & 8.07 & $718\pm34$ & 0.870 & $0.229\pm0.017$ &
$0.447\pm0.021\pm0.034$ &$0.230\pm0.011$ \\
5 & 968.8  & 5.51 & $524\pm34$ & 0.870 & $0.186\pm0.020$ &
$0.589\pm0.039\pm0.065$ &$0.267\pm0.017$ \\
6 & 980.3  & 7.70 & $654\pm37$ & 0.900 & $0.216\pm0.018$ &
$0.436\pm0.025\pm0.038$ &$0.216\pm0.011$ \\
7 & 990.4 & 8.77 & $624\pm38$ & 0.920 & $0.183\pm0.019$ &
$0.422\pm0.026\pm0.045$ &$0.204\pm0.013$ \\
8 & 1003.5  & 20.06 & $1075\pm 50$ & 0.947 & $0.151\pm0.014$ &
$0.374\pm0.018\pm0.035$ &$0.186\pm0.010$ \\
\hline
\end{tabular}
\end{table*}

    The numbers of  found  $n\bar{n}$ events are  listed in the  
Table~\ref{tab:crsect} with the total number close to 6000.
The Table shows only statistical errors of the fitting. 
A source of systematic error  in the $n\bar{n}$ event 
number can be uncertainties in the magnitide and shape of the 
time spectrum of the beam and cosmic background. The error introduced
by these sources is 15 events at  $E_b$=1000 MeV and less than 8
events at lower energies. These values are much lower than
statitistical errors in the Table~\ref{tab:crsect} and are not taken
into account in what follows. 

\begin{figure*}
\includegraphics[width=0.48\textwidth]{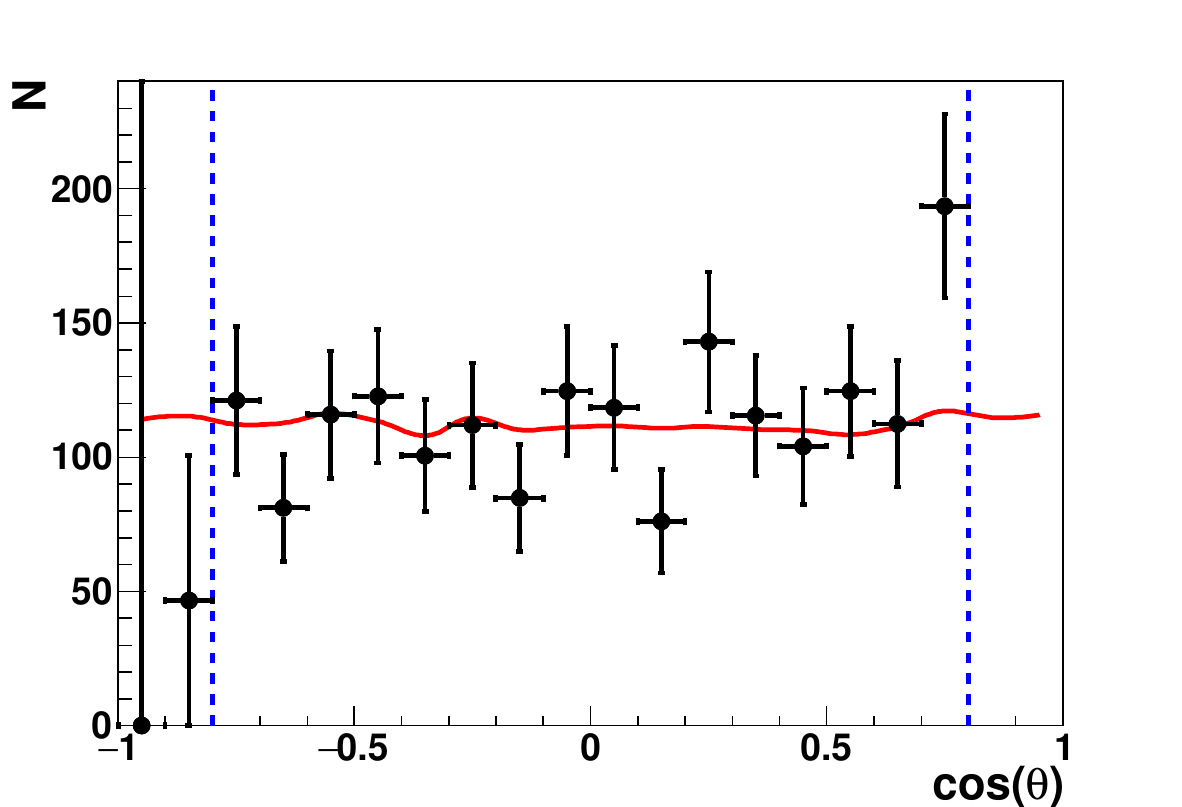} 
\includegraphics[width=0.48\textwidth]{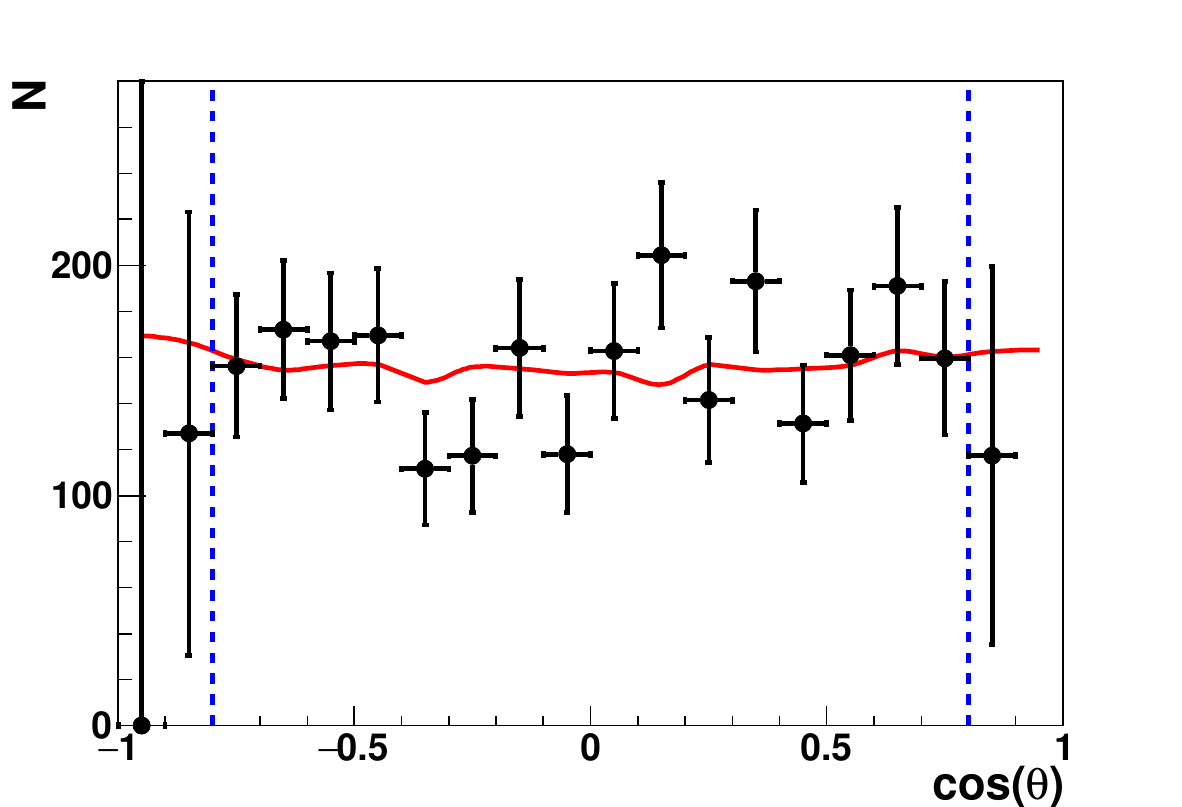}
\caption {The antineutron $\cos\theta_a$ distribution for data 
(points with error bars) and MC (horizontal line) at  
$E_b=970$ MeV (left panel) and $E_b=1000$  MeV (right panel). 
Dotted vertical lines correspond to the polar angle cutoff. 
\label{fig:costh}}
\end{figure*}
%
\begin{figure*}
\includegraphics[width=0.48\textwidth]{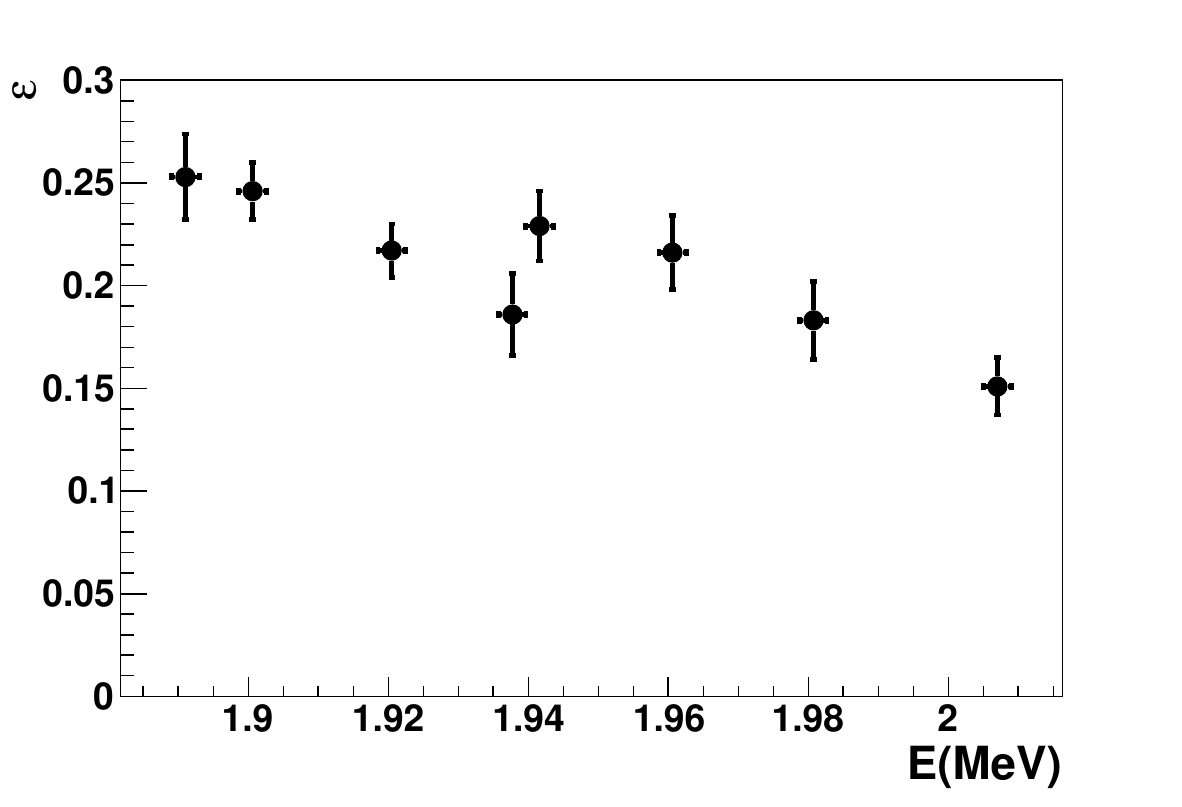} \hfill
\includegraphics[width=0.48\textwidth]{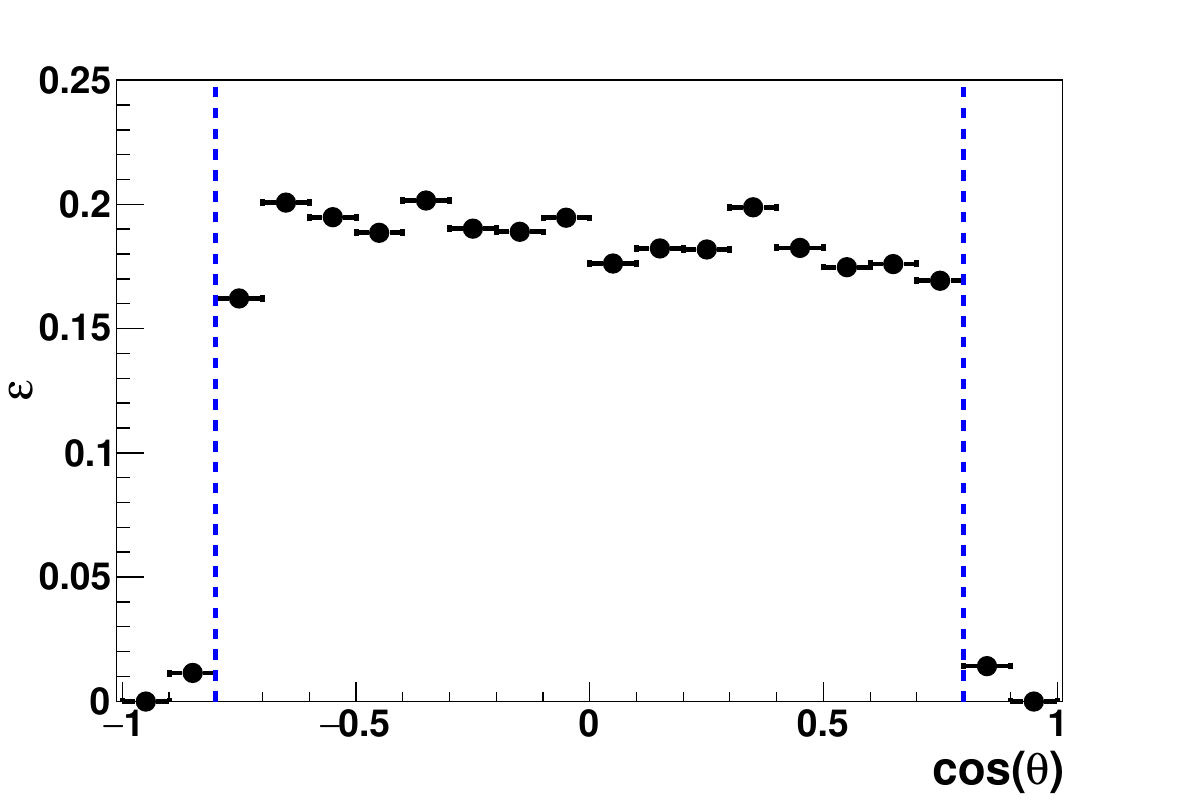} \\ 
\parbox[h]{0.47\textwidth}{\caption { The corrected detection efficiency  
versus energy.}
\label{fig:efenr}} \hfill
\parbox[h]{0.47\textwidth}{\caption{The MC detection efficiency
as a function of antineutron $\cos{\theta}$ at $E_b$=960 MeV.
Dotted vertical lines correspond to the
polar angle cutoff.  }
\label{fig:efth}}
\end{figure*}
\section{Antineutron  angular  distribution\label{sec:cosinth}}
  The antineutron production angle  $\theta_n$  is determined by the 
direction of the event momentum with an accuracy of about 5 degrees. 
Distribution over  $\cos\theta_n$ for data and MC events is shown in 
Fig.~\ref{fig:costh}. The MC simulation was done using Eq.~(\ref{eqB1})
with the assumption $G_E=G_M$. The detection efficiency in the selection 
interval $36^\circ<\theta_n<144^\circ$ is 80\%. It is seen 
from the Fig.~\ref{fig:costh} that the data and MC distributions agree 
well with  each other, which  confirms the MC angular model.   
It is also worth noting that the previous measurements of the  $|G_E/G_M|$ 
value \cite{SNND} also do not contradict the hypothesis $G_E=G_M$. 

\section{Detection efficiency\label{sec:Efficy}}
   The detection efficiency $\varepsilon$ versus energy under accepted  
selection conditions (Section \ref{sec:EvSelect}) is shown in
Fig.~\ref{fig:efenr}. When calculating $\varepsilon$ we used the MC
simulation of the $e^+e^-\to n\bar{n}$ process with the GEANT4
toolkit~\cite{GEANT4}, version 10.5. In addition, the simulation 
included the beam  energy spread $\sim~1$ MeV and the emission of 
photons by initial electrons and positrons. The simulation also took
into account non-operating detector channels as well as overlaps of
the beam background with recorded events. To do this, during the
experiment, with a pulse generator, synchronized with the moment
of beam collision, special superposition events were recorded, which
were subsequently superimposed on MC events.  The detection efficiency 
$\varepsilon$ in Fig.~\ref{fig:efenr} is corrected for the difference 
between the data and MC. This correction is discussed later.
Numerical values of the  efficiency are given in the  
Table~\ref{tab:crsect}. A decrease of $\varepsilon$  with energy can 
be explained by the energy dependence of the selection parameters, 
as well as by an increase in the energy that goes beyond the calorimeter. 
In Fig.~\ref{fig:efth} the angular detection efficiency is shown at
the beam energy $E_b=960$ MeV. 

\begin{figure*}
\includegraphics[width=0.48\textwidth]{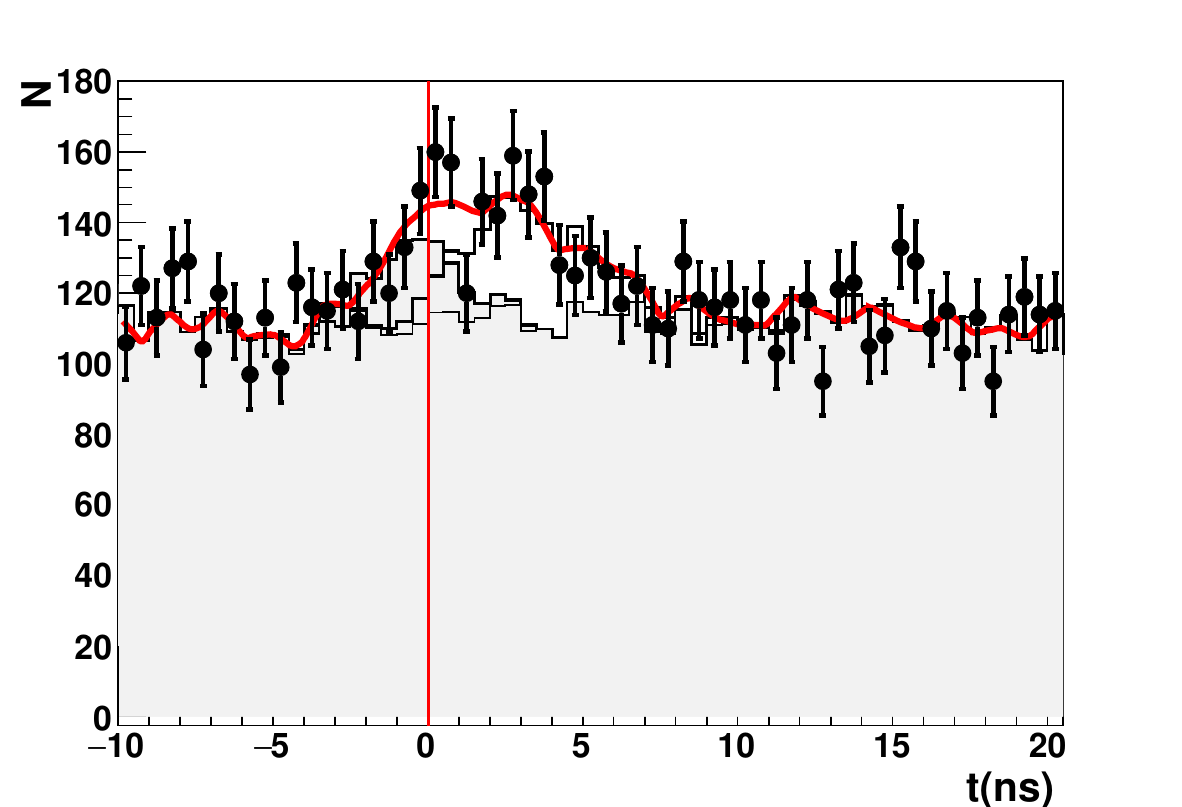} \hfill
\includegraphics[width=0.48\textwidth]{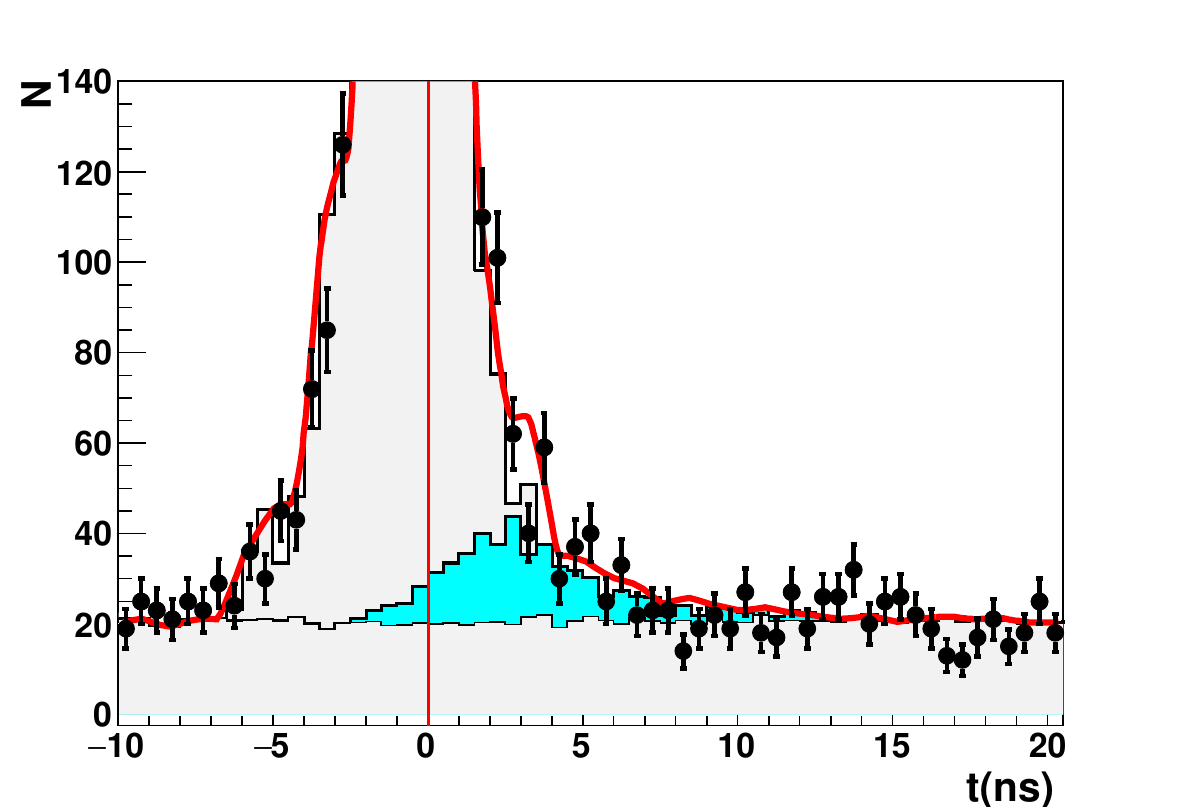} \\
\caption { The event time spectra with inverted selection conditions. 
Left panel --- inverted 2-nd group cut at $E_b=970$ MeV, 
Right panel --- inverted 3-d group cut with $0.7 E_b<E_{\rm cal}<E_b$
at $E_b=960$ MeV. Light shaded histogram shows the background 
distribution. Dark shaded histogram in the right plot is $n\bar{n}$
contribution.
\label{fig:testsel}}
\end{figure*}

The detection efficiency in our measurement is of order of  20\%. 
It is important to find out how correctly the proportion of events
outside the selection condition is simulated. Corrections were
calculated fot three groups of selection conditions described in
chapter \ref{sec:EvSelect}.
To do this, we invert the selection conditions for each 
selection group and then
calculate the corresponding corrections $\delta$ for detection
efficiency in each of 8 energy points as follows:
\begin{equation}
\delta=\frac{n_0}{n_0+n_1}\frac{m_0+m_1}{m_0},
\label{effcor}
\end{equation}
where $n_0$ ($n_1$) is the number of $n\bar{n}$ events determined 
with standard (inverted) selection cuts. These numbers were calculated 
during the time spectra fitting with the Eq.\ref{eq17}, as it is described in
the chapter \ref{sec:Tim19}.  The values $m_0$ and $m_1$ refer respectively 
to the MC simulation event numbers.   Examples of the time spectra 
obtained with inverted conditions are shown in Fig.~\ref{fig:testsel}.

\begin{table*}
\centering
\caption{
The beam energy ($E_b$), the correction to the detection efficiency 
$\delta_1$ from SND internal system,
the correction $\delta_2$ from SND external system,
the correction $\delta_3$ from the EMC thresholds,
the correction $\delta_E$ from the lost EMC energy,
the correction $\delta_t$ is the total correction.
\label{tab:corrs}}
\begin{tabular}{ccccccc}
&$E_b$ (MeV) &$\delta_1$ &$\delta_2$ &$\delta_3$ &$\delta_E$ & $\delta_t$ \\
\hline
1 & 945.5  & $0.991\pm0.022$& $1.292\pm0.092$& $0.971\pm0.038$&
$1.005\pm0.005$  & $1.249\pm0.102$ \\
2 & 950.3  & $0.977\pm0.018$& $1.214\pm0.062$& $0.985\pm0.030$&
$1.009\pm0.009$  &$1.179\pm0.072$ \\
3 & 960.3  & $9.966\pm0.019$& $1.077\pm0.050$& $0.992\pm0.028$&
$1.012\pm0.012$  &$1.044\pm0.062$ \\
4 & 970.8  & $0.949\pm0.021$& $1.198\pm0.061$& $0.980\pm0.050$&
$1.018\pm0.018$  &$1.134\pm0.084$ \\
5 & 968.8  & $0.958\pm0.027$& $1.031\pm0.080$& $0.896\pm0.044$&
$1.018\pm0.018$  &$0.901\pm0.097$ \\
6 & 980.3  & $0.997\pm0.031$& $1.102\pm0.073$& $0.986\pm0.043$&
$1.021\pm0.021$  &$1.106\pm0.093$ \\
7 & 990.4  & $0.925\pm0.033$& $1.131\pm0.080$& $0.889\pm0.041$&
$1.024\pm0.024$  &$0.952\pm0.099$ \\
8 & 1003.5  & $0.915\pm0.024$& $1.065\pm0.056$& $0.796\pm0.028$&
$1.028\pm0.028$  &$0.797\pm0.073$ \\
\hline
\end{tabular}
\end{table*}

  The first group of selection conditions includes the requirement of
no charged tracks in an event. When studying inverse selections we
assume the presence of central charged tracks with $D_{xy}>0.5$ cm,
where $D_{xy}$ is the distance between the track and the axis of the beams.  
A possible background from the  related process $e^+e^-\to p\bar{p}$
should be discussed here. In the energy region $E_b > 960$ MeV protons
and antiproptons give central collinear  tracks and rejected by the requirement
$D_{xy}>0.5$ cm, as well as events of other processes of $e^+e^-$
annihilation with charged tracks. However at $E_b < 960$ MeV the protons
and antiproptons are slow and stop at the collider vacuum pipe. 
In this case the antiproton annihilates  with the production  of charged
tracks, wchich can be with $D_{xy}>0.5$ cm. But here, too, the
$e^+e^-\to p\bar{p}$ background is suppressed by the fitting of time
spectrum, since the annihilation delay time does not exceed 1 ns.
For the second group the inverted selection conditions were used
without changes. For the third group of selection conditions, 
a partial inversion was used, 
that is, the condition $0.7 E_b<E_{\rm cal}<E_b$ was applied.

   An additional correction arises from the events, in which the
antineutrons pass the calorimeter without interaction, and from the
events with a small calorimeter energy $E_{\rm cal}<0.7 E_b$. These
events are not taken into analysis due to the large background and
therefore not available for correction with the procedure described
above. Their share in MC varies from 1.9\% at the energy $E_b$=945 
MeV to 8.5\% at $E_b$=1000 MeV. It was previously noted in  
chapter ~\ref{sec:Tim19},  that to desribe the shape of data time spectrum 
the contribution of the process of antineutron scattering in MC  
should be reduced by a factor of 1.5. With such a change, the proportion  
of events with  $E_{\rm cal}<0.7 E_b$ in MC reduces to 1.4\% at
$E_b$=945 to 5.7\% at  $E_b$=1000 MeV. The difference between these 
values is taken into account as an additional correction $\delta_E$ to 
the detection efficiency with the   100\% of uncertainty.

  The measured by selection groups corrections 
$\delta_1, \delta_2, \delta_3$, as well as  $ \delta_E$, are multiplied
$\delta_t = \delta_1 \delta_2 \delta_3 \delta_E$ and all are given 
in the Table~\ref{tab:corrs}.  
It can be seen, that the total efficiency correction $\delta_t$  changes 
in limits 0.8---1.25 with energy, what is explained by the strong
energy dependence of the antineutron absobtion length. 

   The corrected detection efficiency is obtained from the MC efficiency
by multiplying by the total correction $\delta_t$.    
The values of the corrected efficiency are given along with systematic 
errors in the Table~\ref{tab:crsect}. Here, unlke our previous
measurement \cite{SNND}, the corrections in different energy points are not
correlated.

\section{The measured $e^+e^-\to n\bar{n}$ cross section \label{sec:Crosct}}
\begin{figure*}[h]
\includegraphics[width=0.49\textwidth]{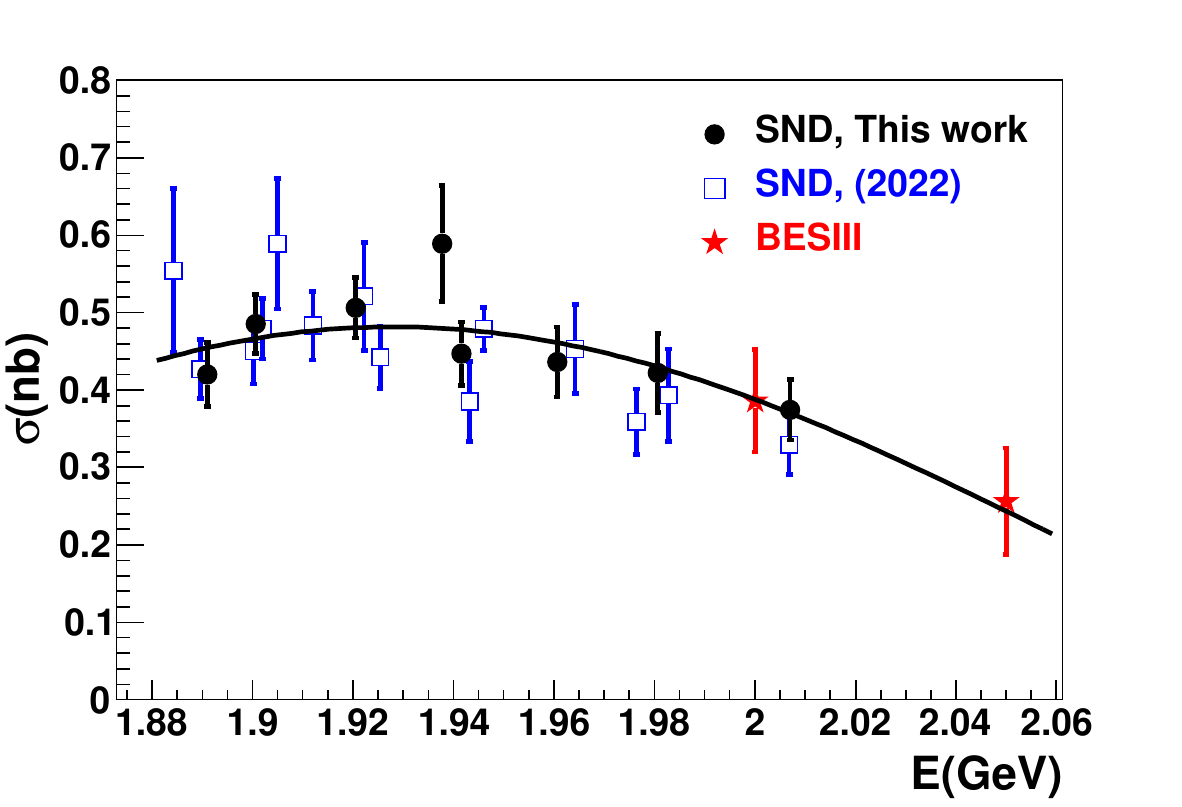} \hfill
\includegraphics[width=0.49\textwidth]{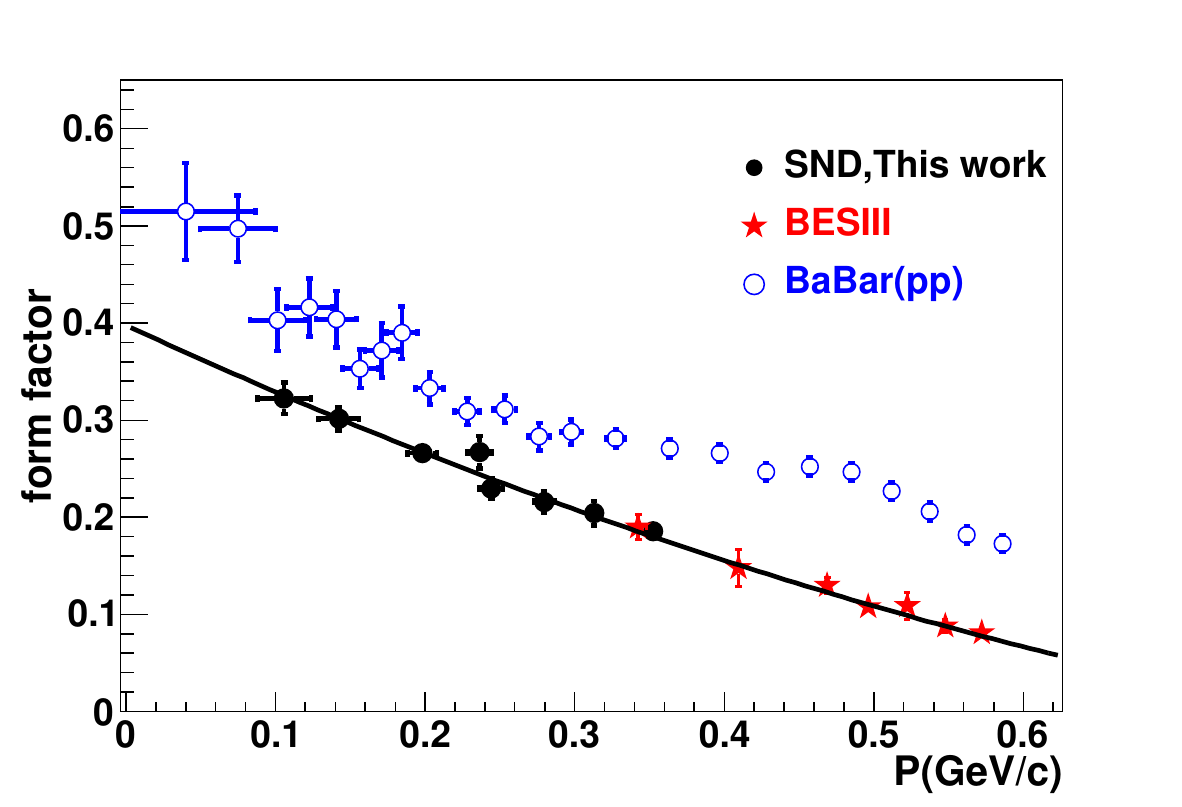} \\
\parbox[h]{0.45\textwidth}{
\caption {The $e^+e^-\to n\bar{n}$ cross section measured
in this work in comparison with previous SND measurement \cite{SNND}
and BESIII~\cite{BES} data.  The solid curve is the fit result. 
Only new SND and BESIII data were used in the fit.}
\label{fig:csect1}} \hfill
\parbox[h]{0.45\textwidth}{
\caption {The measured neutron timelike formfactor (solid black points)
compared with BESIII~\cite{BES} (red stars) and 
BABAR~\cite{Babar} proton formfactor (blue empty squares)
The solid line is the polinomial fit, described in the text.}
\label{fig:ffnn}}
\end{figure*}
Using the number of $n\bar{n}$ events $N_{n\bar{n}}$ , 
luminosity $L$ and detection
efficiency $\varepsilon$ (Table \ref{tab:crsect}), the
visible cross section  $\sigma_{vis}(E)= N_{n\bar{n}}/L\varepsilon$   
can be calculated. The Born cross section $\sigma(E)$ we need is
related to the visible cross  $\sigma_{vis}(E)$ in the following form :
\begin{eqnarray}
\sigma_{vis}(E)&=&\sigma(E)(1+\delta(E))\nonumber\\
&=&\int_{-\infty}^{+\infty}G(E^\prime,E)dE^\prime\nonumber\\
&&\int_0^{x_{max}}W(s,x)\sigma(s(1-x))dx,
\label{eqB4}
\end{eqnarray}
where  $W(s,x)$ is the radiator function~\cite{Radcor} describing
emission of  photons with energy $xE_b$ by initial electrons and positrons, 
$G(E^\prime,E)$ is a Gaussian function describing the c.m. energy spread.
In  function $W(s,x)$ the 
contribution of the vacuum polarization is not taken  into account,  
so our Born cross section is  a ``dressed'' cross section.
The factor $(1+\delta(E))$ takes into account both the radiative
corrections and beam energy spread. 
This factor is calculated in each of 8 energy points using the Born cross
section, obtained by the fitting of the visible cross section using 
Eq.~\ref{eqB4}. The energy dependence of the Born cross section   
is described by Eq.\ref{eqB2}, in which the neutron effective formfactor
has a form of a second order polinomial function, as shown in more
detail in the next chapter.

The measured Born cross section is shown in the Fig.~\ref{fig:csect1}
and listed in the Table~\ref{tab:crsect}. The dominant contribution 
into systematic error is made by the detection efficiency
correction error, given in the Table \ref{tab:corrs}.
Uncertainties in the value of luminosity (1\%) and radiative correction
(2\%) are also taken into account. In Fig.~\ref{fig:csect1} the total
statistical and systematic error is shown. In comparison with our
preceding work ~\cite{SNND}, the measured cross section has 2 times
lower statistical error and 1.5 times lower systematic error. 
At the maximum energy $E=2$ GeV our cross section 
is in good agreement with the last  BESIII  measurement~\cite{BES}.

\section{The neutron effective timelike formfactor \label{sec:fform} }

The effective neutron form factor calculated from the measured cross 
section using Eq.~(\ref{eqB2}) is listed in the Table \ref{tab:crsect}
and shown in Fig.~\ref{fig:ffnn} as a function of the antineutron momentum 
together with the BESIII data~\cite{BES} and the proton
effective form factor measured by the BABAR experiment~\cite{Babar}.
The curve in Fig.~\ref{fig:ffnn}, approximating the form factor, 
is a second order polinimial $|F_n|=a_0+a_1p_n+a_2p_n^2$, 
in which the parameters $a_i$ are obtained  during fitting and $p_n$ 
is the antineutron momentum. The following parameters values are
obtained: $a_0=0.398\pm0.022$, $a_1=-0.713\pm0.126$, $a_2=0.268\pm0.166$.
When the fitting curve continues to zero momentum, the expected value of 
the neutron formfactor at the threshold will be $a_0\simeq$0.4. One
can see from the Fig.~\ref{fig:ffnn}, that the proton formfactor
noticably larger that the neutron one and their ratio near the
threshold could be close to $3/2$.

\section{Summary \label{sec:summary} }
The experiment to measure the $e^+e^-\to n\bar{n}$ cross section 
and the neutron timelike form factor has been
carried out with the SND detector at the VEPP-2000 $e^+e^-$ collider
in the energy region from 1891 to 2007 MeV. 
The measured  $e^+e^-\to n\bar{n}$ cross varies  with energy 
within 0.4$\div$0.6 nb 
and agrees  with recent  SND measurement SND~\cite{SNND}, however  
has 2 times better statistical accuracy.  At the maximum energy our cross
section  is in agreement   with the last BESIII measurement~\cite{BES}.
The  neutron effective timelike form factor  is extracted from the
measured cross section using Eq.\ref{eqB2}.  Form factor decreases 
with energy from 0.3 to 0.2.  In the value, the  neutron form factor 
turns out to be noticeably less than the proton one.

{\bf ACKNOWLEDGMENTS}. This work was carried out on the RSF fund grant
No. 23-22-00011.

\end{document}